\documentclass[twocolumn,prl,superscriptaddress]{revtex4} 
\usepackage{epsfig}
\usepackage{graphicx}
\usepackage{amsmath}
\usepackage{color,colordvi,graphicx,pstcol}

\newcommand{\ket}[1]{|#1\rangle}

\newcommand{\hide}[1]{}

\newcommand{\eq}[1]{Eq.\,(\ref{#1})}

\newcommand{\fig}[1]{Fig.\,\ref{#1}}

\newcommand{\ra}{\ensuremath{\rightarrow}}
\newcommand{\lra}{\ensuremath{\longrightarrow}}

\begin{document}

\title{Schemes for robust quantum computation with polar molecules
}
\author{S. F. Yelin}
\affiliation{Department of Physics, University of Connecticut,
Storrs, CT 06269} 
\affiliation{ITAMP, Harvard-Smithsonian Center
for Astrophysics, Cambridge, MA 02138}
\author{K. Kirby}
\affiliation{ITAMP, Harvard-Smithsonian Center
for Astrophysics, Cambridge, MA 02138}
\author{Robin C\^ot\'e}
\affiliation{Department of Physics, University of Connecticut,
Storrs, CT 06269} 
\affiliation{ITAMP, Harvard-Smithsonian Center
for Astrophysics, Cambridge, MA 02138}
\date{\today}

\begin{abstract}
We propose to use a new platform -- ultracold polar molecules -- 
for quantum computing with switchable interactions. The on/off
switch is accomplished by selective excitation of one of the $\ket{0}$ 
or $\ket{1}$  qubits -- long-lived molecular states -- to an ``excited'' 
molecular state with a considerably different dipole moment. We 
describe various schemes based on this switching of dipolar interactions 
where the selective excitation between ground and excited states is 
accomplished via optical, micro-wave, or electric fields. 
We also generalize the schemes to take advantage of the {\it dipole 
blockade} mechanism when dipolar interactions are very strong. These
schemes can be realized in several recently proposed architectures.
\end{abstract}

\maketitle

Quantum computing is one of the most rapidly developing areas in physics
today. For certain tasks, quantum computers have 
significant potential to outperform classical computers \cite{roadmap}.  
Several platforms are being investigated to implement these ideas,
{\it e.g.}, using atomic, molecular and optical, 
condensed matter, and other systems. A key challenge in all 
of these approaches is to identify {\it strong} and 
{\it controllable} interactions that would allow for the 
creation of fast quantum operations with minimal decoherence.

Quantum information processing
makes use of quantum superposition in which
the fundamental piece of information, called a {\it qubit}, consists
of a superposition of quantum states, denoted $\ket{0}$ and
$\ket{1}$.  The building blocks of a quantum computer consist
of ``gate'' operations, in which a coherent change in the state
of one qubit can be brought about through a carefully controlled
interaction with another qubit, and the result is dependent on
the state of the second qubit. In order to implement reversible quantum logic
operations it is essential to address these quantum states
coherently. 

Of the various platforms proposed to implement quantum computers,
trapped ions and neutral atoms are especially attractive \cite{roadmap}.
Trapped ions \cite{ions}
exhibit strong interactions and are relatively easy to control, while
neutral atoms \cite{atoms} have long coherence times and techniques 
to cool and trap them are well developed.  
Polar molecules represent a new platform that might incorporate 
the biggest advantages of both atoms and ions and even bridge the 
gap with condensed matter physics approaches ({\it e.g.} 
molecule-chips \cite{mol-chip} or microtraps connected 
to superconducting wires \cite{mol-super}). 
They have long coherence times like  neutral atoms, and strong 
interactions like trapped ions. However, contrary to ions, 
the interactions can be made ``switchable,'' a feature which 
would help to simplify phase gates and minimize decoherence.
Advances in cooling \cite{mol-cool} and storing \cite{mol-trap} 
techniques for molecules are beginning to make possible 
the required accurate manipulation of single molecules.

In this Letter, we investigate the implementation of universal two-qubit
logic gates in realistic systems, using ultracold polar molecules. 
As opposed to other schemes using polar molecules \cite{demille-qc},
such as vibrational eigenstates \cite{qc-v} and optimal
control \cite{kosloff-qc}, our approach is based on the ability
to ``switch'' on and off dipole interactions between polar molecules. 
For this, we use the fact that all heteronuclear molecules have (a) 
different dipole moments depending on their electronic, 
vibrational, or rotational states, and (b) zero expectation 
value for the dipole moment in the N=0 rotational state.

\begin{figure}[htb]
\includegraphics[width=0.7\linewidth]{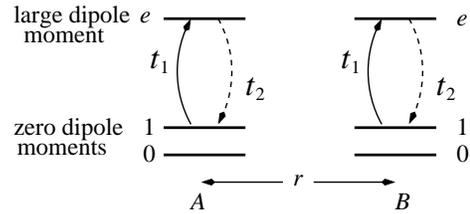}
 \caption{\protect\label{fig:setup} Phase gate: two 
  molecules $A$ and $B$ separated by $r$ are prepared in a
  superposition of states $\ket{0}$ and $\ket{1}$. At
  $t_1=0$, we excite $|1\rangle$ of both into $|e\rangle$:
  both interact via dipole-dipole interactions, and acquire 
  a phase $\phi$. At time $t_2=\tau$ such that
  $\phi =\pi$, we stimulate coherently both $\ket{e}$
  back to $\ket{1}$.}
\end{figure}
We first describe the generic setup to obtain a phase gate, or universal 
two-qubit operation, in \fig{fig:setup}. We assume that the molecules are 
individually addressable by optical or microwave fields, and choose 
$\ket{0}$ and $\ket{1}$ as, for example, hyperfine states, in
part of a zero-dipole-moment manifold 
in a level with a long coherence time  
and $\ket{e}$ 
is a metastable state in a large-dipole-moment manifold. 
Single-qubit rotations can be accomplished with optical 
or microwave fields. The initial states of two individual 
sites $A$ and $B$ can be prepared in a superposition state, 
e.g. using $\pi/2$ Raman pulses. A one- or two-photon transition 
couples $\ket{1}$ and $\ket{e}$ coherently, 
but not $\ket{0}$ and $\ket{e}$. This can always be accomplished by 
either polarization or frequency selection. The molecules interact 
via a dipole-dipole interaction only if both are in the $\ket{e}$ state, 
and acquire a phase $\phi (t)$. After a time $t=\tau$ such that 
$\phi =\pi$, we coherently stimulate the states $\ket{e}$ back 
to $|1\rangle$. This can be summarized by
\[
\label{eq:phasegate}
\begin{array}{ccccrcr}
\ket{00} & \stackrel{\tiny\pi-{\rm pulse}}{\longrightarrow} 
         & \ket{00} & \stackrel{\tiny\rm dip-dip}{\longrightarrow}
         & \ket{00} & \stackrel{\tiny\pi-{\rm pulse}}{\longrightarrow} 
         & \ket{00}\\
\ket{01} & \longrightarrow & \ket{0e} & \longrightarrow 
         & \ket{0e} & \longrightarrow & \ket{01}\\
\ket{10} & \longrightarrow & \ket{e0} & \longrightarrow 
         & \ket{e0} & \longrightarrow & \ket{10}\\
\ket{11} & \longrightarrow & \ket{ee} & \longrightarrow 
         & -\ket{ee} & \longrightarrow & -\ket{11}
\end{array}.
\]
The resulting transformation corresponds to a phase gate. 
We desire the wave function, expressed as $e^{-iE\Delta t/\hbar}$,
to acquire a phase shift of $\pi$, thus becoming $e^{-i\pi}$, as
A and B experience dipole-dipole interaction in state $\ket{e}$.
The $\pi$-phase shift produced in the time $\tau$ between 
the exciting and de-exciting $\pi$-pulses is given by
\begin{equation}
  \label{eq:dip-dip}
  \phi=\pi=\frac{1}{\hbar}\int_0^\tau d\tau'\frac{d^2}{r^3}
             \left(3\cos^2\theta-1\right) \rho_e^2(\tau'),
\end{equation}
where $d$ and $\rho_e$ are the dipole moment and fractional population 
in the excited state, $r$ the distance between 
molecules $A$ and $B$, and $\theta$ the angle 
between the dipole moments. This formulation allows for finite 
excitation and de-excitation times and imperfect $\pi$-pulses.

\begin{figure}[h]
\includegraphics[width=0.65\linewidth]{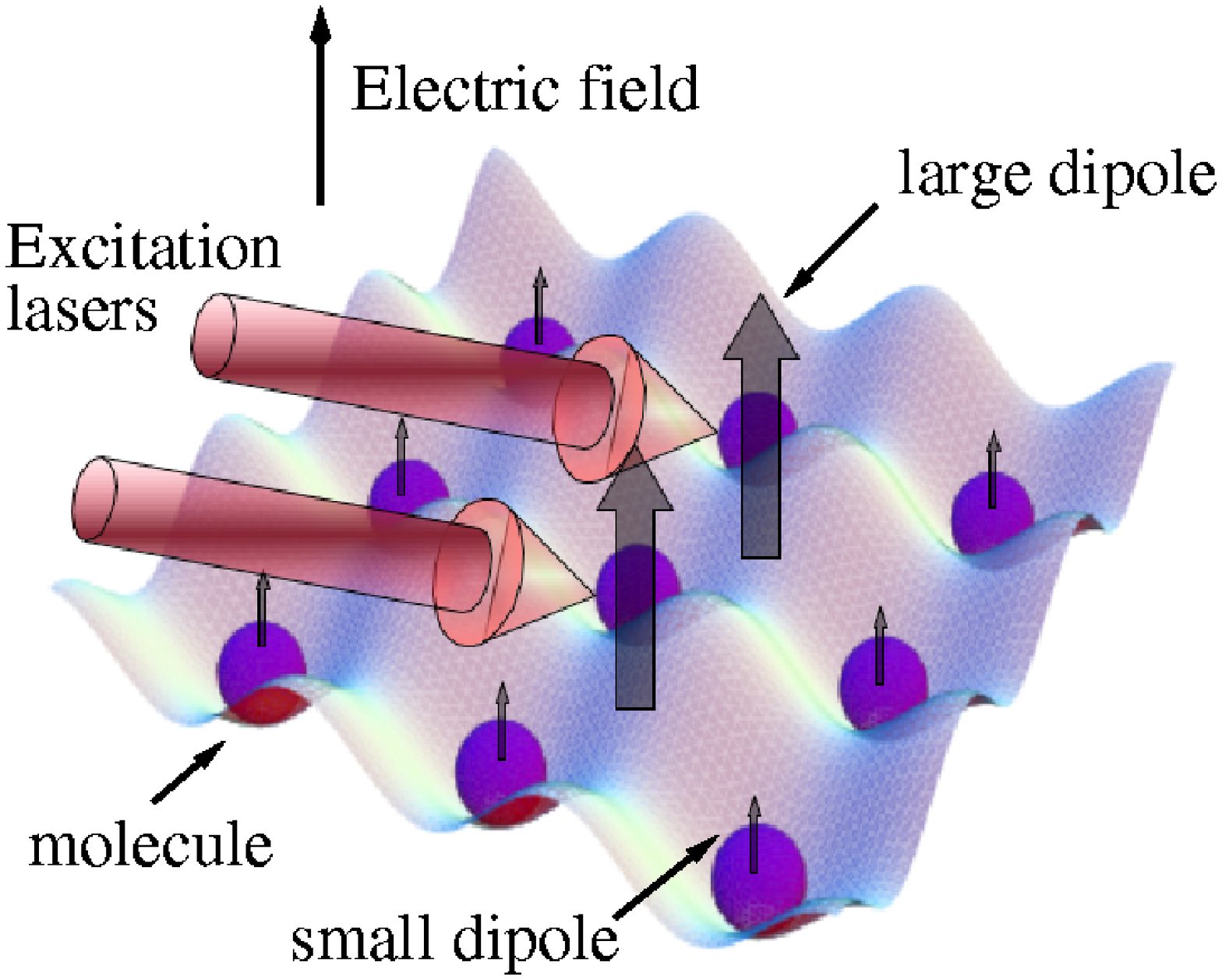}\\[.5cm]
\includegraphics[width=0.65\linewidth]{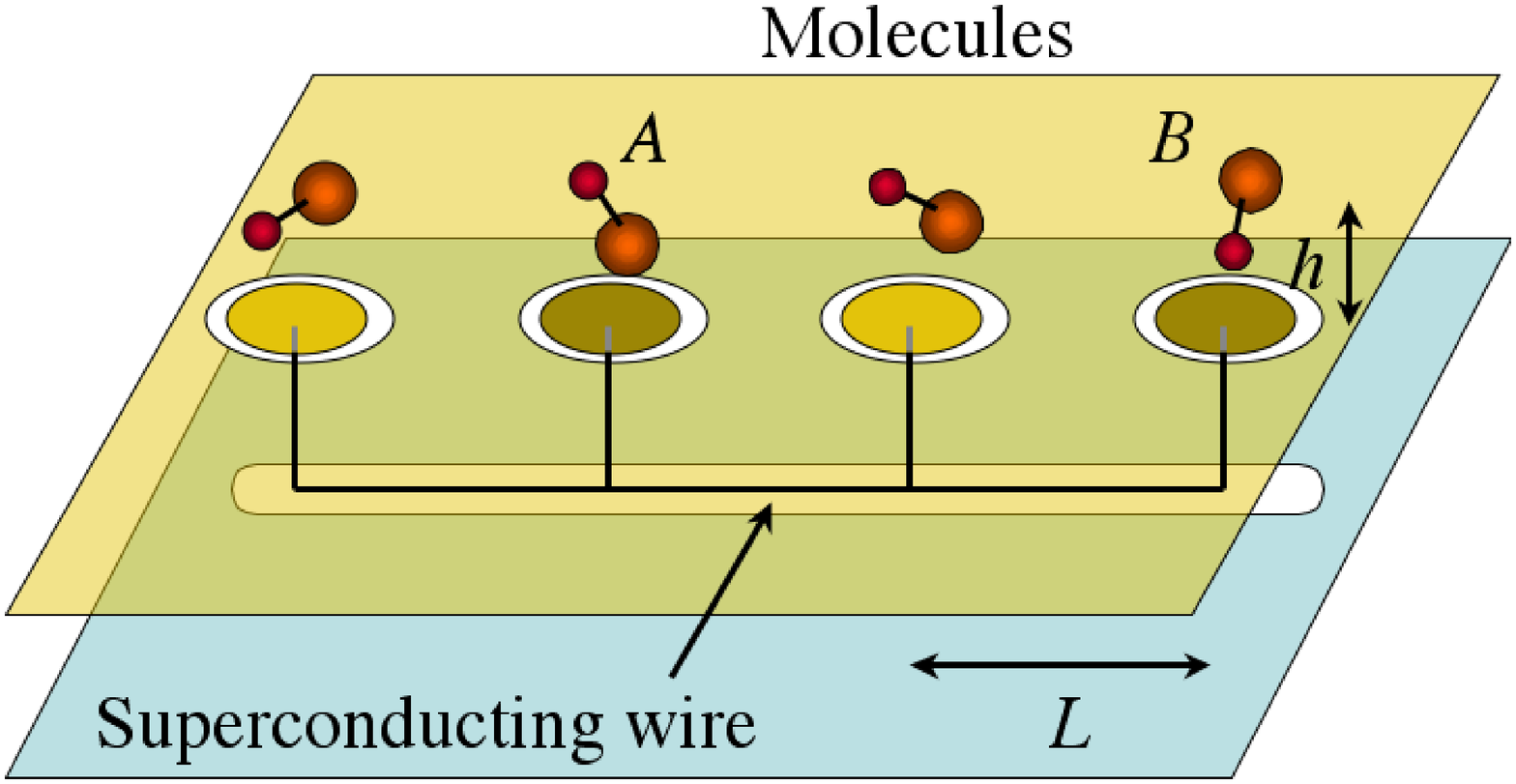}
\caption{\protect\label{fig:array}
Setups: (Top) molecules individually addressable by lasers
are stored in an optical lattice, (Bottom) superconducting 
wires are used to ``deliver'' the interaction. In both, 
molecules are selectively excited, and interact only 
if both are in $\ket{e}$. }
\end{figure}

To implement the scheme, molecules are stored in a 1D or 2D
array so that their dipole moments could be aligned by an 
electric field perpendicular to the array. We assume the 
full development of the storage and addressing capabilities 
of two recently proposed architectures (see Fig.~\ref{fig:array}). 
The first is an optical lattice with a lattice spacing of 
about 1 $\mu$m, as suggested by DeMille \cite{demille-qc}. 
Using a DC field for dipole alignment during trapping naturally 
allows the repulsive dipole-dipole interaction to aide with 
homogeneous distribution in the lattice. In this case, addressing
single qubits can be accomplished by either using the 
inhomogeneous DC electric fields proposed by DeMille to 
create individualized transition frequencies, or by individual 
addressing with light in the visible part of the frequency spectrum. 
The second  architecture is based on a ``strip wire'' architecture, 
as suggested at Yale and Harvard \cite{mol-super}; molecules sit on 
their own small microwave traps which also serve for addressing, and 
are connected via a superconducting wire that allows for long-range 
dipole-dipole interaction, effectively replacing the $1/r^3$ 
term in \eq{eq:dip-dip} with $1/h^2r$, where $h$ is the distance 
of the molecules from the wire. Here, all the fields need to be in 
the microwave range.  

We now describe three possible setups utilizing variations of
our switchable phase-gate scheme. The first system is based on 
carbon monoxide (CO). As far as dipolar molecules are concerned, 
CO is an anomaly; while its electronic ground state $X^1\Sigma^+$ 
has a very small dipole moment ($\mu\approx 0.1$ D in the vibrational ground state 
which is expected to be the easiest to trap), there exists a 
very long-lived ($\tau_{\rm life}\approx 10-1000$ ms) excited 
electronic state $a^3\Pi$ with a large dipole moment, 
$\mu\approx 1.5$ D. As ``0'' and ``1,'' we choose, for example, 
two hyperfine states of $X^1\Sigma^+, v=0, N=0$ of 
$^{13}$CO \cite{footCO}.
With a hyperfine splitting of about 1 MHz, selective excitation from 
$\ket{1}$ to $\ket{e}$ is possible. The transition frequency 
between $X^1\Sigma^+$ and $a^3\Pi$ is in the UV  
(about 48,000 cm$^{-1}$), and the optical lattice 
architecture would be the ideal choice. With a coherence 
time in an optical lattice of a few seconds 
\cite{demille-qc} and possibly much less \cite{footres} and 
a necessary dipole-dipole interaction time of several 
milliseconds, there can be about $10^3$ operations. 
The scheme, however, is very straightforward, the techniques are
in place or nearly so, and CO is a very well studied 
molecule \cite{klemperer-meijer}.

A more common situation can be found in molecules such as
alkali hydrides or mixed alkali dimers, e.g. LiH or LiCs. 
These molecules have large permanent dipole moments $\mu$ 
(as large as 7 D) in their ground electronic state $X^1\Sigma^+$ 
(for $\ket{0}$ and $\ket{1}$), and a metastable electronic 
state $a^3\Sigma^+$ (for $\ket{e}$) for which
the potential well is located at large nuclear separation and 
supports at least one bound state; in most cases, these triplet 
states have permanent dipole moments close to zero. These 
properties can be used to implement a scheme in all important 
points similar to the CO scheme, except for three details. 
First, the phase gate would be ``inverted'', i.e. 
$\ket{00}\ra -\ket{00}$, $\ket{01}\ra\ket{01}$, 
$\ket{10}\ra\ket{10}$, and $\ket{11}\ra\ket{11}$. Second, it  
requires the molecules to be stored, with the help of an 
aligning DC electric field, in the large-dipole state which 
would most likely lead to seriously shortened coherence times. 
In addition, the interaction would happen for {\em all} molecules, 
not just the two we wish to be coupled by a phase gate. However, 
this can be mitigated by switching on an aligning DC field only 
during interaction times, and for exactly a 2$\pi$ phase shift.

For any molecule in a pure $N=0$ rotational state, 
the expectation value of its dipole moment
 is zero. Such states can acquire a dipole moment 
by the application of an electric field that mixes $N=0$ 
and $N=1$ states. So, by adding 
together the $2\pi$ phase shift using a DC field and the 
``negative'' $\pi$ phase shift for the molecules in the 
$\ket{e}$ state, the phase gate is given by
\[
\label{eq:phasegate_inverted}
\begin{array}{rcrcrcrcrcr}
\ket{00} & \stackrel{\tiny{\rm exc + DC}}{\lra} & \ket{00}
         & \stackrel{\tiny\pi}{\lra} & -\ket{00} 
         & \stackrel{\tiny{\rm de-exc}}{\lra} & -\ket{00} 
         & \stackrel{\tiny DC}{\lra} & \ket{00}\\
\ket{01} & \lra & \ket{0e} &\lra& \ket{0e} &\lra& \ket{01} 
         & \lra& -\ket{01}\\
\ket{10} & \lra & \ket{e0} &\lra& \ket{e0} &\lra& \ket{10} 
         & \lra& -\ket{10}\\
\ket{11} & \lra & \ket{ee} &\lra& \ket{ee} &\lra& \ket{11} 
         & \lra& -\ket{11}
\end{array}.
\]
Note that the scheme described for CO 
could be adapted for these molecules by using 
two different vibrational states of  $a^3\Sigma^+$ as 
$\ket{0}$ and $\ket{1}$, and a low-level vibrational 
state of $X^1\Sigma^+$ as $\ket{e}$. 

\begin{figure}[ht]
\includegraphics[width=0.8\linewidth]{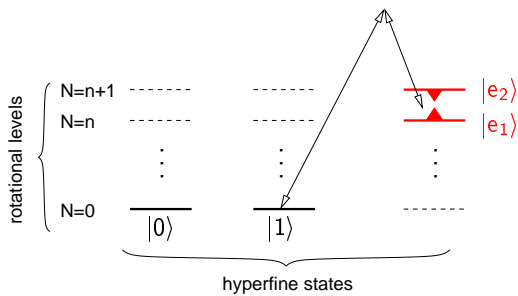}
\caption{\protect\label{fig:rotscheme}
	Example of level system for ``rotational scheme'': all states are part of the electronic and vibrational ground state. $\ket{0}$ and $\ket{1}$ are in different hyperfine states. $\ket{0}, \ket{1}$ are $\ket{N=0}$ states, $\ket{e}\propto\ket{e_1}+\ket{e_2}$ is a superposition between two adjacent rotational states $\ket{N=n}$ and $\ket{N=n+1}$.}
\end{figure}

The last setup we propose here is the ``rotational scheme''.
It utilizes the fact that in the rotational ground state $N=0$ the dipole moment $\mu$ is, in fact, zero. We choose for all states the electronic and vibrational ground state. While $\ket{0}, \ket{1}$ are also in the rotational ground state $N=0$, $\ket{e}$ is the {\em superposition} of neighboring rotational states $\ket{e}=\ket{e_1}+\ket{e_2}$, as shown in \fig{fig:rotscheme} \footnote{$\ket{e}$ is subject to the following restrictions: (1) $\ket{e}$ has to be coupled by a two- or more photon transition to $\ket{1}$ and this photon combination must not couple $\ket{0}$; (2) in order to have a non-zero dipole moment there has to be an electric dipole allowed transition between $\ket{e_1}$ and $\ket{e_2}$; (3) there must not be an electric dipole allowed transition between $\ket{e_{1/2}}$ and either $\ket{0}$ or $\ket{1}$, since otherwise spontaneous diffusion of the excitation in the molecular sample could occur.}.  Because both $\ket{0}$ and $\ket{1}$ are in the absolute ground state with exactly zero dipole moment, this system has several advantages: maximum coherence time, ease of storage, and no residual dipole-dipole interaction. Moreover, any polar molecule can be used with this scheme, as long as it has at least two hyperfine states. One interesting choice would be NaCl with a dipole moment of up to 10 D.

Given the fact that rotational levels are spaced in the GHz range and thus only low-frequency photons are required, this scheme is suitable for both the optical lattice and the superconducting wire 
architectures; for a dipole moment $\mu=10$ D, $r=10$ $\mu$m, and 
$h=0.1$ $\mu$m, the necessary interaction time is of the order of 3 $\mu$s. 
With a coherence time of the order of 100 ms - 1 s, this setup would thus
allow for $10^5-10^6$ operations. 

If the sites can be addressed individually and the dipole-dipole
interactions are very strong, the previous schemes could take 
advantage of the so-called dipole {\em blockade} mechanism. 
This mechanism has been introduced for quantum information
processing with Rydberg atoms in \cite{jaksch}, and generalized 
to mesoscopic ensembles \cite{lukin}. A variant of the dipole 
blockade based on the strong van der Waals interactions between 
Rydberg atoms, known as the {\it vdW blockade}, has been explored 
experimentally \cite{prl-tong}. The underlying principle goes as 
follows: strong Rydberg-Rydberg interactions shift the energy 
levels, so that one atom can be resonantly excited into a Rydberg 
state, but additional Rydberg excitations are prevented by the 
large shifts. 
\begin{figure}[htb]
  \includegraphics[width=0.8\linewidth]{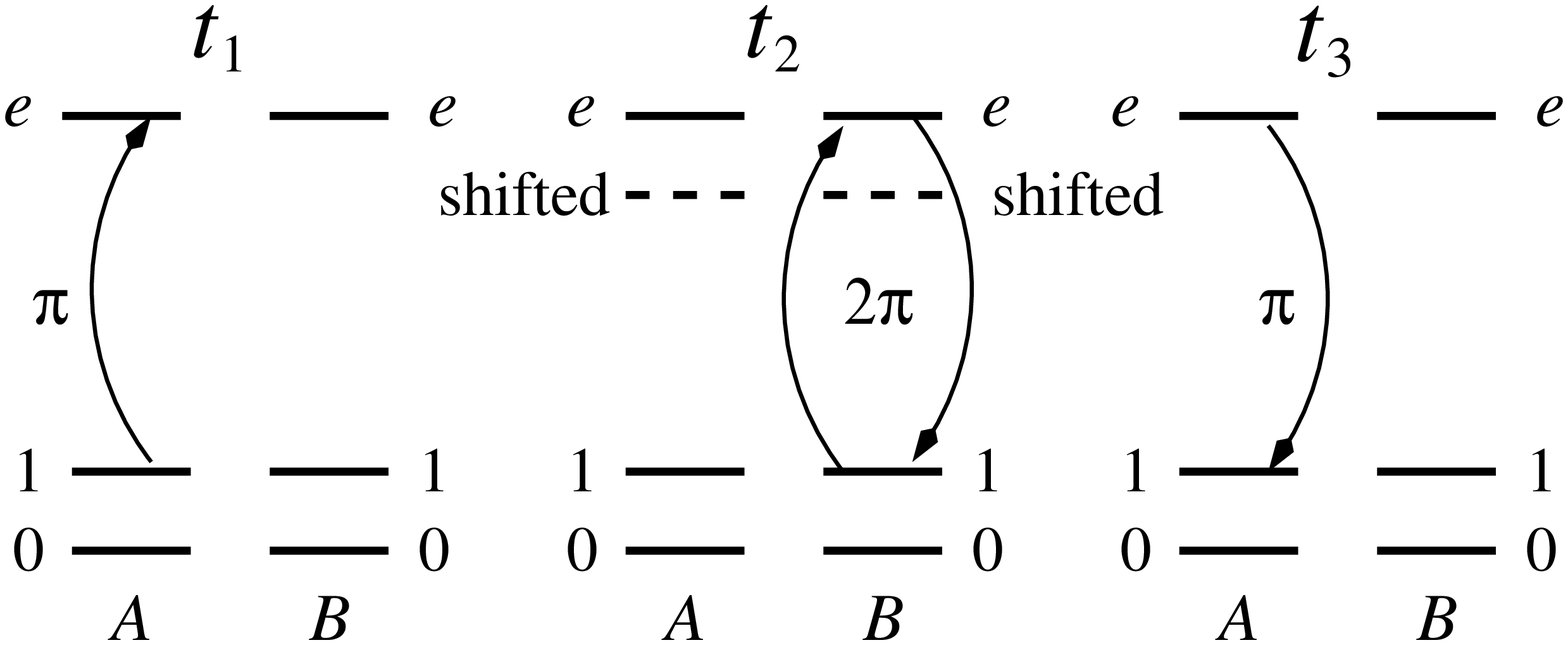}
  \caption{
            Principle of the dipole blockade (see text).}
   \label{fig:dip_bloc}
\end{figure}

The same idea can be applied to polar molecules. If the
dipole-dipole interaction is strong enough, {\i.e.} larger than
the bandwidth of the excitation laser, the doubly-excited state
corresponding to $\ket{ee}$ will be shifted out of resonance
and never excited. If both sites $A$ and $B$ are addressable 
individually, the ability to drive a $2\pi$ transition in 
site $B$ depends on whether site $A$ is excited 
(see \fig{fig:dip_bloc}). At $t_{1}$, we apply a $\pi$ pulse 
to molecule $A$ and populate the state $\ket{e}$. At $t_{2}$ 
we apply a second pulse $(2\pi )$ to molecule $B$: if $A$ 
is already in $|e\rangle_{A}$, the dipole-dipole interaction 
shifts the state $|e\rangle_{B}$, the photon is off-resonance, 
hence no transition. If $A$ is not in $|e\rangle_{A}$, 
$B$ acquires a phase of $\pi$ after the process. At $t_{3}$,
we de-excite $A$ with another $\pi$ pulse; in summary 
\[
\label{eq:dip-scheme}
\begin{array}{rcrcrcr}
&t_1& &t_2& &t_3& \\
\ket{00} &  & \ket{00} &  & \ket{00} &  & \ket{00}\\
\ket{01} &  & \ket{01} & \longrightarrow & -\ket{01} &  & -\ket{01}\\
\ket{10} & \longrightarrow & i\ket{e0} & & i\ket{e0} 
         & \longrightarrow & -\ket{10}\\
\ket{11} & \longrightarrow & i\ket{e1} 
         & \longrightarrow\hspace*{-.5cm}{\sf\bf x} & i\ket{e1} 
         & \longrightarrow & -\ket{11}
\end{array}.
\]
This scheme is robust with respect to the separation between 
the molecules; as long as the excitation is blockaded, 
the exact separation is not important. 

A key operation at the end of several qubit operations 
is the readout of the quantum registers. Several approaches
could be employed with polar molecules. For example, selective 
ionization of one of the states (0 or 1) and the detection of 
molecular ions can be readily accomplished. However, this is
a destructive method, since the molecule is lost after the readout, 
and the site would need to be refilled. A different method uses
a ``cycling'' fluorescent transition in which the molecules decay 
after irradiation directly back into the state from which they came.
Although this might be more difficult for molecules than for atoms 
because of the large number of molecular levels, it offers the 
advantage of being ``non-destructive.'' Another approach based on 
recent work on evanescent-wave mirrors for polar molecules might 
yield promising results \cite{mol-mirror}; while ``0'' would 
stick to the wall, ``1'' could be reflected. Because reflection takes 
place far away from the surface of the mirror, it might help 
to minimize decoherence due to shorter range interactions with 
the surface. The schemes described above may suffer from various
sources of error. As our schemes rely only on internal molecular 
states, there is a possibility that some of the molecules may be 
translationally ``hot.'' If molecules are not in the motional 
ground state of the trap, there can be considerable uncertainty 
and variation in the separation between molecules, which can affect 
the exact phase. For example, during a $\sim 1$ $\mu$s gate time, 
the motion of RbCs molecules at 10 $\mu$K can lead to $\sim 3$\% 
variation in the phase. We can control and reduce such 
error, e.g., using molecules with larger dipole moments, larger separations, 
shorter gate times, or lower temperatures. Note that decoherence and 
uncertainty due to molecular motion can be completely eliminated 
using {\it dipole blockade}, leading to higher fidelity \cite{lukin}.

Finally, several technical issues may affect the implementation
of our schemes, such as turning on and off electric fields, 
misalignments of dipoles, decoherence in an optical lattice 
(e.g., incoherent photon scattering or ionization), or imperfect 
excitation pulses; these can be overcome, e.g., by trapping molecules 
with evanescent-wave mirrors, using Stimulated Raman Adiabatic 
Passage (STIRAP) or chirped pulses. Other physical 
effects such as DC and AC Stark mixing, spontaneous 
decay of metastable molecular states, or rovibrational quenching
if sites contain more than one molecule, can be avoided by a 
judicious choice of molecules and states, and careful loading
of sites.

In summary, we propose a new platform that combines the advantages of
both neutral atoms, such as long coherence times, and trapped ions,
such as strong interactions. Contrary to ions, the interactions
can be made ``switchable,'' a feature helping to simplify phase 
gates considerably and thus to minimize decoherence. Using these 
techniques, up to 10$^6$ operations should be obtainable in the 
available coherence time. Finally, the possibility exists to 
bridge the gap with condensed matter devices, using polar 
molecules instead of quantum dots in circuits with superconducting 
wires that convey the interaction.

We would like to thank D. DeMille, J. Doyle, M. Lukin, and D. Petrov for
fruitful discussions. Partial funding from the National Science 
Foundation and the Research Corporation is gratefully acknowledged.

%
%


\begin{thebibliography}{99}
\bibitem{footCO} 
   Note that, since $^{12}$CO has a zero nuclear spin,
   isotopic $^{13}$CO would have to be used.
\bibitem{footres} 
   The residual dipole moment in the ground state may lead to 
   much more decoherence than is given by the lattice.
\bibitem{note-rotation-scheme}
    It is important to ensure that no single-photon 
    transitions are possible between $\ket{0}$ or 
    $\ket{1}$ and $\ket{e}$. Otherwise a virtual 
    photon exchange could happen between the ``active'' 
    qubits (in $\ket{e}$) and any neighboring molecule, resulting in 
    ``diffusion'' of the excitation.
\end{thebibliography}
\end{document}